\documentclass[letterpaper, 10 pt, conference]{ieeeconf}  

\IEEEoverridecommandlockouts                              
\usepackage{amsmath}
\usepackage{tikz}
\usepackage{amssymb}
\usepackage{multirow}

 \newcommand{\qed}{\hfill\blacksquare}

\title{\LARGE \bf
Dissipativity Conditions for Maximum Dynamic Loadability}

\author{\thanks{This material is based upon work supported by the National Science Foundation Graduate Research Fellowship Program under Grant No. 2141064 and  the NSF Grant No. ECCS-2002570 entitled ASCENT: Boosting Cyber and Physical Resilience of Power Electronics-Dominated Distribution Grids.} Riley E. Lawson$^1$\thanks{$^1$ Riley E. Lawson is a Ph.D. candidate at the Laboratory for Information and Decisition Systems (LIDS) of Massachusetts Institute of Technology, Cambridge, MA, USA {\tt lawsonri@mit.edu}} and Marija D. Ilic$^2$\thanks{$^2$ Marija D. Ilic is a Joint EECS Adjunct Porfessor and a Senior Research Scientist at Laboratory for Information and Decision Systems (LIDS), Massachusetts Institute of Technology (MIT), Cambridge, USA. She is also  a Professor Emerita in the Electrical and Computer Engineering (ECE) Department of Carnegie Mellon University, Pittsburgh, PA. {\tt ilic@mit.edu}}}

\begin{document}

\maketitle
\thispagestyle{empty}
\pagestyle{empty}

\begin{abstract}
In this paper we consider a possibility of stabilizing very fast electromagnetic interactions between Inverter Based Resources (IBRs), known as the Control Induced System Stability problems. We propose that when these oscillatory interactions are controlled the ability of the grid to deliver power to loads at high rates will be greatly increased. We refer to this grid property as the dynamic grid loadability. The approach is to start by modeling the dynamical behavior of all components.  Next, to avoid excessive complexity, interactions between components are captured in terms of unified technology-agnostic aggregate variables, instantaneous power and rate of change of instantaneous reactive  power. Sufficient dissipativity conditions in terms of rate of change of energy conversion in components themselves and bounds on their rate of change of interactions  are derived in support of achieving  the  maximum system loadability. These physically intuitive conditions are then used to derive methods to increase loadability using high switching frequency reactive power sources. Numerical simulations confirm the theoretical calculations, and shows dynamic load-side reactive power support increases stable dynamic loadability regions.

\end{abstract}

\section{INTRODUCTION}
Growing interest in decarbonization efforts have led to large scale introductions of IBRs at every level of power system operation. Transitioning from conventional generators to IBR generation introduces new challenges for system operators. One key challenge is adapting conventional methods of analyzing system stability for the changing portfolio of resources used in power generation, as the time scales at which conventional and new resources act are very different \cite{siljak2002robust}, \cite{green2018modelling}. 
\par This paper is motivated by the need to model these dynamics with a particular objective of controlling very fast electromagnetic interactions between IBRs, known as the Control Induced System Stability problems \cite{ercot}. We propose that when these oscillations are controlled the ability of the grid to deliver power to loads at high rates will be greatly increased. We refer to this grid property as the dynamic grid loadability. To avoid confusion, we differentiate between the concept of system loadability and long studied maximum power transfer conditions, and their meaning in the static and dynamic cases. Recall that the maximum power transfer was conceptualized in circuit theory, and is achieved through impedance matching between the source and load \cite{desoer1973maximum}. In \cite{calvaer1982maximum} it was shown in power systems attempting to apply this notion of maximum power transfer results in operation outside allowable ranges, and is thus constrained by operational requirements. In this early work, maximum power transfer is viewed as the maximum possible power a load may consume, regardless of system stability. Conversely, system loadability considers stability. Dynamic loadability considers the dynamics of components, and has no direct notion of an underlying equilibria of the system, but instead gives limits on the rates at which the load power can be served. 
\par Investigating exactly how these additional dynamics impact dynamic loadability is essential. Early work on investigating component dynamics found regions of feasible equilibria, commonly referred to as PV curves, may become unstable when considering the dynamics of conventional generators, though model the lines and loads through algebraic relationships \cite{sauer1993maximum}. More recently, for high frequency components such as IBRs it has been shown it is insufficient to only consider the dynamics of the source, but that the stability of the inverter itself requires dynamic models of transmission lines connected to the resources as well \cite{gross2019effect}, \cite{mohammed2023impacts}.
\par Analyzing the dynamics of every component is made even more challenging by the complex internal behavior of each component.  Properly analyzing and controlling the dynamic behavior of these fast-acting IBRs requires enhancing  modeling of components \cite{lacerda2023phasor}. In this paper we propose to use modeling which directly captures interactions of aggregate component dynamics, which may be represented in a unified way \cite{ilic2018multi}, \cite{jaddivada2021feasible}. Use of aggregate energy dynamics of components  makes it possible to represent interactions within the interconnected system in a unified manner by defining  rates of change of shared variables \cite{willems2007behavioral} in terms of  instantaneous power and rate of change of reactive power \cite{ilic2018multi}. This fundamental interpretation of shared variables in physical systems  in terms of rates of change of power is independent from the type of component internal energy conversion process.  
\par As of yet, available literature has not analyzed the question of system loadability where every component in the system is treated in the dynamical sense. This paper aims to utilize these unified modeling techniques to study the problem of maximum dynamic loadability in this way. The contributions of this work are two fold. First, we show explicitly how the unified modeling approach of \cite{ilic2018multi} can be used to study system level properties of transmission networks. Second, we demonstrate methods of improving maximum dynamic loadability through fast-acting reactive power compensation. 
\par The remainder of this paper is organized as follows. The need for modeling RLC dynamics of transmission line dynamics is conceptualized in Section \ref{timescalesection}. The general structure of energy dynamics modeling is presented in Section \ref{energydynamics} for completeness. This modeling structure is applied to a general RLC transmission network, sufficient conditions on dynamic maximum loadability are derived, and a method for increasing system stability and loadability are shown in Section \ref{dissipativity}. The theoretical conditions are validated using a numerical example on a simple system in Section \ref{simresultssection}. Finally, concluding thoughts are provided in Section \ref{conclusion}.

\section{THE NEED FOR MODELING RLC DYNAMICS OF TRANSMISSION LINES}
\label{timescalesection}
As discussed above, the time scales at which conventional generators and IBRs operate are widely separated. We present here a demonstration of the breakdown of the assumption that transmission line behavior is nearly instantaneous, as is typically taken in today's electric power system studies, when source dynamics are much faster. 
\par
\begin{figure}
    \centering
    \includegraphics[width = \columnwidth]{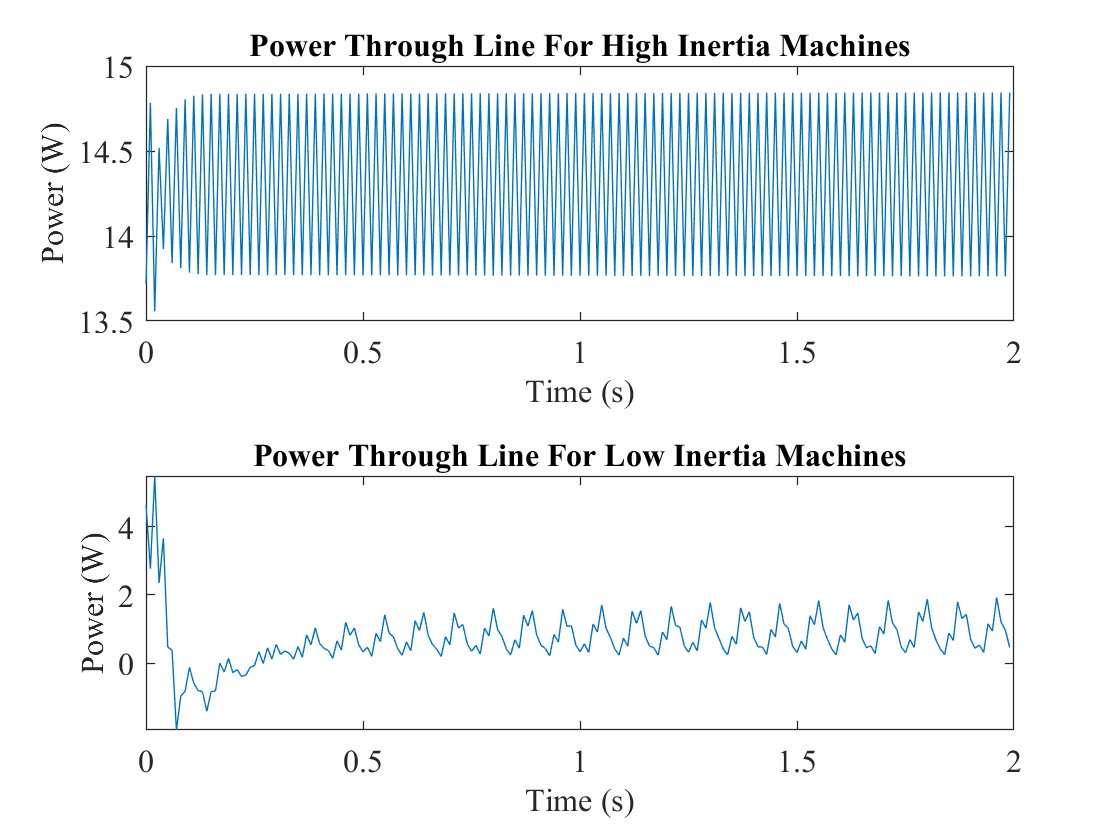}
    \caption{Instantaneous  power dynamics  through a transmission line connecting two synchronous machines modeled using a Type $IV_1$ model \cite{iliczab}. The top plot shows the dynamics for two machines with very high inertia, and the bottom plot for those with very low inertia.}
    \label{diffFreqResponse}
\end{figure}

 Consider a simple system comprised of two synchronous machines connected by a transmission line, shown in Fig. \ref{fig:GTLL}. To analyze the effects of reactive component dynamics on the response of the system, the transmission line is modeled as a lumped parameter equivalent $\pi$-model, including losses. The system is simulated over a short duration under two different conditions. First, the generators are taken to have very high inertial constants ($J=100$), emblematic of large conventional generators. Second, the generators are taken to have very small inertial constants ($J=1$), representative of systems with higher penetrations of IBRs. All other generator parameters are identical. Fig. \ref{diffFreqResponse} shows the response of this system under these two generator configurations.
\par In the high inertia case the changes in the machine outputs are slower relative to the line time constant. Therefore, the line internal dynamics settle before the machine outputs change, and the variation in power through the line is at the system frequency, 60 Hz. Conversely, in the low inertia case, the machine outputs change before the line internal dynamics settle, and thus the interconnected components interact, creating new dynamical phenomena from the superposition of the two.

\section{ENERGY DYNAMICS MODELING} \label{energydynamics}
Due to the broad variety of technologies serving the functions of generation, transmission, and loads, it is desirable to express the interactions between these components in a generalizable, technologically agnostic way. Consider some dynamical component, governed by the system in (\ref{generaldynsys}), where $x$ is the vector of system states, $u$ the control inputs, $m$ the exogenous inputs, and $r$ the port inputs.
\begin{subequations}
\label{generaldynsys}
    \begin{align}    
        \dot{x} &= f(x, u, m, r), \ x(0) = x_0  \\
        y &= g(x, u, m, r)
    \end{align}
\end{subequations}
It was shown in \cite{ilic2018multi}, the generic system (\ref{generaldynsys}) may be transformed into the interactive, linear system of (\ref{energydynamsys}), which is in terms of internal energy storage, and interactions with the rest of the system through its ports.
\begin{subequations}    
    \label{energydynamsys}
    \begin{align}
        \dot{E} &= P - \frac{E}{\tau} = p\\
        \dot{p} &= 4E_t - \dot{Q}
    \end{align}
\end{subequations}
Given some inertia matrix, $H(x)$ describing the devices internal components, which may be dependent on the current states, the energy stored in the component, $E$ is given by (\ref{storedE}). Conversely, given some damping matrix, $B(x)$, the dissipation of a component is found as (\ref{dissipatedE}). 
\begin{subequations}
    \begin{align}
    \label{storedE}
        E(x) &= \frac{1}{2} x^TH(x)x \\
        \label{dissipatedE}
        D(x) &= \frac{1}{2} x^TB(x)x
    \end{align}
\end{subequations}
The time constant is the ratio of stored energy to dissipation \cite{hill1994aperture}
\begin{equation}
\label{timeConstant}
    \tau = \frac{E(x)}{D(x)}
\end{equation}
The energy in tangent space is defined as the energy function evaluated on the tangent space of the state variables
\begin{equation}
\label{tangentEnergy}
    E_t = \frac{1}{2} \dot{x}^T H(x) \dot{x}
\end{equation}
\par $P$ and $\dot{Q}$ describe the components interactions with the rest of the system at its ports. Using the multi-energy analogies presented in \cite{jeltsema2009multidomain}, $P$ and $\dot{Q}$ are defined based on port variables as in (\ref{PQdot}) \cite{wyatt1990time}.
\begin{subequations}
\label{PQdot}
    \begin{align}
        P &= e^T f \\
        \dot{Q} &= e^T \frac{df}{dt} - f^T \frac{de}{dt}
    \end{align}
\end{subequations}
The notion of energy in tangent space was recently employed by \cite{forni2013differential}, \cite{kosaraju2017control} to derive notions of differential passivity, considering derivatives of port variables in addition to the port variables themselves.
If the component in question has more than one port, the terms for $P$ and $\dot{Q}$ can be replaced with the summation of the individual values calculated at each port of the component.
\par \noindent \textbf{Lemma 1.} \textit{The quantities $P$, $\dot{P}$, and $\dot{Q}$ are Tellegen quantities.}
\par \textbf{Proof:} $P$ being a Tellegen quantity follows from Tellgen's theorem. Consider the sum and difference form of Tellegen's theorem from \cite{penfield1970generalized}, given in (\ref{sumdifftellegen}).
\begin{equation}
\label{sumdifftellegen}
    \sum_\alpha \Lambda ' i_\alpha \Lambda '' v_\alpha \pm \Lambda '' i_\alpha \Lambda ' v_\alpha =  \sum_p \Lambda ' i_p \Lambda '' v_p \pm \Lambda '' i_p \Lambda ' v_p
\end{equation}
Taking as Kirchoff operators $\Lambda' = \frac{d}{dt}$ and $\Lambda'' = I$, the identity operator, $\dot{P}$ and $\dot{Q}$ being Tellegen quantities follows directly. $\qed$

\section{DISSIPATIVITY CONDITIONS FOR MAXIMUM DYNAMIC LOADABILITY} \label{dissipativity}
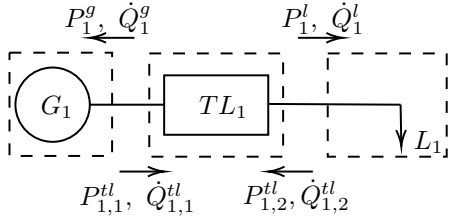
\begin{figure}
    \centering
    
    \tikzset{every picture/.style={line width=0.75pt}} 
    
    \begin{tikzpicture}[x=0.75pt,y=0.75pt,yscale=-.75,xscale=.75]
    
    \draw   (110,115) .. controls (110,101.19) and (121.19,90) .. (135,90) .. controls (148.81,90) and (160,101.19) .. (160,115) .. controls (160,128.81) and (148.81,140) .. (135,140) .. controls (121.19,140) and (110,128.81) .. (110,115) -- cycle ;
    \draw    (160,115) -- (210,115) ;
    \draw   (210,95.25) -- (280,95.25) -- (280,135.25) -- (210,135.25) -- cycle ;
    \draw    (280.5,114.5) -- (305,114.64) -- (369,115) ;
    \draw    (369,115) -- (369.46,141) ;
    \draw [shift={(369.5,143)}, rotate = 268.98] [color={rgb, 255:red, 0; green, 0; blue, 0 }  ][line width=0.75]    (10.93,-3.29) .. controls (6.95,-1.4) and (3.31,-0.3) .. (0,0) .. controls (3.31,0.3) and (6.95,1.4) .. (10.93,3.29)   ;
    \draw  [dash pattern={on 4.5pt off 4.5pt}] (106,80) -- (175,80) -- (175,149.5) -- (106,149.5) -- cycle ;
    \draw  [dash pattern={on 4.5pt off 4.5pt}] (200,80.5) -- (290,80.5) -- (290,150) -- (200,150) -- cycle ;
    \draw  [dash pattern={on 4.5pt off 4.5pt}] (320,80.5) -- (400,80.5) -- (400,150) -- (320,150) -- cycle ;
    \draw    (190,70) -- (162,70) ;
    \draw [shift={(160,70)}, rotate = 360] [color={rgb, 255:red, 0; green, 0; blue, 0 }  ][line width=0.75]    (10.93,-3.29) .. controls (6.95,-1.4) and (3.31,-0.3) .. (0,0) .. controls (3.31,0.3) and (6.95,1.4) .. (10.93,3.29)   ;
    \draw    (180,160) -- (208,160) ;
    \draw [shift={(210,160)}, rotate = 180] [color={rgb, 255:red, 0; green, 0; blue, 0 }  ][line width=0.75]    (10.93,-3.29) .. controls (6.95,-1.4) and (3.31,-0.3) .. (0,0) .. controls (3.31,0.3) and (6.95,1.4) .. (10.93,3.29)   ;
    \draw    (310,160) -- (282,160) ;
    \draw [shift={(280,160)}, rotate = 360] [color={rgb, 255:red, 0; green, 0; blue, 0 }  ][line width=0.75]    (10.93,-3.29) .. controls (6.95,-1.4) and (3.31,-0.3) .. (0,0) .. controls (3.31,0.3) and (6.95,1.4) .. (10.93,3.29)   ;
    \draw    (300,70) -- (328,70) ;
    \draw [shift={(330,70)}, rotate = 180] [color={rgb, 255:red, 0; green, 0; blue, 0 }  ][line width=0.75]    (10.93,-3.29) .. controls (6.95,-1.4) and (3.31,-0.3) .. (0,0) .. controls (3.31,0.3) and (6.95,1.4) .. (10.93,3.29)   ;
    
    \draw (125,106.9) node [anchor=north west][inner sep=0.75pt]    {$G_{1}$};
    \draw (231.5,106.9) node [anchor=north west][inner sep=0.75pt]    {$TL_{1}$};
    \draw (376,130.9) node [anchor=north west][inner sep=0.75pt]    {$L_{1}$};
    \draw (141,44.4) node [anchor=north west][inner sep=0.75pt]    {$P_{1}^{g} ,\ \dot{Q}_{1}^{g}$};
    \draw (151,164.4) node [anchor=north west][inner sep=0.75pt]    {$P_{1,1}^{tl} ,\ \dot{Q}_{1,1}^{tl}$};
    \draw (261,163.4) node [anchor=north west][inner sep=0.75pt]    {$P_{1,2}^{tl} ,\dot{Q}{_{1,2}^{tl}}$};
    \draw (287,44.4) node [anchor=north west][inner sep=0.75pt]    {$P_{1}^{l} ,\ \dot{Q}_{1}^{l}$};
    
    \end{tikzpicture}
    \caption{Simple power transfer system comprised of a single generator $G_1$, transmission line $TL_1$ and load $L_1$, with their associated port interactions noted.}
    \label{fig:GTLL}
\end{figure}

Consider a system comprised of generation sources, transmission lines, and loads. Sources and loads are treated as single port elements, and transmission lines are treated as two port elements. Denote by $G$ the set of all generators, $TL$ the set of all transmission lines, and $L$ the set of all loads. Further, let superscript $g$ denote quantities relating to elements of $G$, $tl$ denote quantities relating to elements of $TL$, and $l$ denote quantities relating to elements of $L$. Further, because transmission lines are two port elements, an additional subscript of 1 denotes quantities on the sending end port of the line, and an additional subscript 2 denotes quantities on the receiving end port. Fig. \ref{fig:GTLL} shows a system comprised of a single generator, line, and load.
\par We aim to establish dissipativity conditions on the system, and use them to determine sufficient conditions on system loadability. Consider the energy in tangent space of the transmission network, representing the potential of the components to do real work. Take as the storage function the system (\ref{storageFunction}), expressing the total energy in tangent space over all components.
\begin{equation}
\label{storageFunction}
    S(t) = 4\int \sum_{i \in TL} E_{t, i}^{tl} dt
\end{equation}
Power is injected into the transmission network at its interfaces with the generating units and loads. Take as the supply function of the system (\ref{supplyFunction}).
\begin{equation}
    \label{supplyFunction}
    s(t) = \sum_{j \in G} \frac{-P_j^g}{\tau_j^g} + \sum_{k \in L} \frac{-P_k^l}{\tau_k^l}
\end{equation}
The negative sign on each injection is used to represent injections into the network, as the sign convention is taken as positive into the port of the component. Further, each injection is scaled by the time constant of the component, accounting for the different time scales at which the components act. 

\noindent \textbf{Theorem 1.} \textit{A transmission network, $TL$, comprised of 2-port transmission lines in an arbitrary configuration, connected at its boundaries to some number of generators, $G$, and loads, $L$, with storage function (\ref{storageFunction}) is dissipative with respect to the supply function (\ref{supplyFunction}) if the total load in the system satisfies:}
\begin{align}
\nonumber
    \sum_{k \in L} &\left( \frac{1}{\tau^{tl}} +\frac{1}{\tau_k^l} \right) P_k^l \leq -\sum_{i \in TL} \frac{{E}_i^{tl}}{(\tau^{tl})^2} + \sum_{k \in L} \left( \dot{P}_k^l + \dot{Q}_k^l \right) \\ 
    \label{powerTransBound} 
    &+ \sum_{j \in G} \left( \dot{P}_j^g + \dot{Q}_j^g - \left(\frac{1}{\tau^{tl}} + \frac{1}{\tau_j^g} \right) P_j^g\right)
\end{align}

\par \textbf{Proof:} Dissipativity requires that
\begin{equation}
\label{dissipativityCond}
    \dot{S}(t) \leq s(t)
\end{equation}
Taking the time derivative of (\ref{storageFunction}), and applying the constituent relationships from (\ref{energydynamsys}) yields
\begin{align}
\nonumber
    \dot{S}(t) &= 4\sum_{i \in TL} E_{t,i}^{tl} \\ \nonumber
    &= \sum_{i \in TL} \left( \dot{p}_i^{tl} + \dot{Q}_{i,1}^{tl} + \dot{Q}_{i,2}^{tl} \right) \\ \label{Sdot1}
    &= \sum_{i \in TL} \left( \frac{-\dot{E}_{i}^{tl}}{\tau_{i}^{tl}}+ \dot{P}_{i,1}^{tl} + \dot{P}_{i,2}^{tl} + \dot{Q}_{i,1}^{tl} + \dot{Q}_{i,2}^{tl} \right)
\end{align}
As was established in Lemma (1), $\dot{P}$ and $\dot{Q}$ are Tellegen quantities. Internal connections in the transmission network will ensure that for any component $i$ there exists some other collection of components $\alpha_1,...,\alpha_n$ connected to the same bus as component $i$ such that $\dot{P}_{i,2}^{tl} + \sum_{r \in \{1,...,n\}} \dot{P}_{\alpha_r, 1}^{tl} = 0$. An identical argument applies to the cancellation of $\dot{Q}$ for internal connections. 
\par For connections at the boundary of the transmission network, lines will be connected to either a load or generator. Consider a boundary connection to a generator. By Lemma 1, there must exist some collection of components $\beta_1,...,\beta_m$ such that $\sum_{r \in \{1,...,m\}} \dot{P}_{\beta_r, 2}^{tl} = \dot{P}^g$. An identical argument applies to interfaces with loads and for the $\dot{Q}$ terms. From this, (\ref{Sdot1}) may be rewritten as
\begin{align}
    &\dot{S}(t) = \sum_{i \in TL} \frac{-\dot{E}_i^{tl}}{\tau_{i}^{tl}} - \sum_{j \in G} \left( \dot{P}_j^g + \dot{Q}_j^g \right) - \sum_{k \in l} \left(\dot{P}_k^l + \dot{Q}_k^l \right) \\
    &= \sum_{i \in TL} \frac{{E}_i^{tl}-P_i^{tl}\tau_i^{tl}}{(\tau_{i}^{tl})^2} - \sum_{j \in G} \left( \dot{P}_j^g + \dot{Q}_j^g \right) - \sum_{k \in l} \left(\dot{P}_k^l + \dot{Q}_k^l \right)
\end{align}
\par
Again using Lemma 1, given that $P_i^{tl}$ are also Tellegen quantities, identical arguments as above allow for the cancellation of all terms internal to the transmission network, and leave only the injections from the interfaces. Further, we may take as an approximation that $\tau_i^{tl} = \tau^{tl}$ $\forall i \in TL$, as for similar voltage levels, the $r/l$ ratio of transmission lines are nearly constant. 
\begin{align}
    \nonumber
    \dot{S}(t) = \sum_{i \in TL} \frac{{E}_i^{tl}}{(\tau^{tl})^2} &- \sum_{j \in G} \left( \dot{P}_j^g + \dot{Q}_j^g - \frac{P_j^g}{\tau^{tl}} \right) \\ \label{Sdotfinal}
    &- \sum_{k \in l} \left(\dot{P}_k^l + \dot{Q}_k^l - \frac{P_k^l}{\tau^{tl}}\right) 
\end{align}
Applying the dissipativity condition (\ref{dissipativityCond}) to (\ref{Sdotfinal}) allows the system load to be bounded as in (\ref{powerTransBound}), dependent on the energy change in the transmission system divided by the line time constant, the rates of change of the source and load power injections, and the actual power injections from sources. $\qed$

Examining the conditions in (\ref{powerTransBound}), the right hand side of the bound has four key components. Each component relates to a rate of change of power in the device, as it is a bound on the dynamic loadability. The first term in the right hand side is the energy in transmission network components divided by their time constants squared. This corresponds to the rate of change of power being processed by the corresponding transmission component. 
\par Recall the used sign convention dictates positive as into the component. Two of the terms relate to the rate at which the power and reactive power being respectively generated and demanded by the sources and loads are changing. Finally is the term relating to the rate at which power is transmitted out of the sources, which is limited both by its internal dynamics and those of the transmission network. 


\subsection{Methods for Increasing Dynamic Loadability}
The upper bound in (\ref{powerTransBound}) suggests methods by which the total system loadability may be increased. If the load increases at some rate, the conditions may imply an equal and opposite rate of change at a generator would balance this change. This would, however, have impacts on the internal energy storage dynamics of the transmission network, as the energies injected at the two ends would propagate through the network. These changes to the energies stored in the transmission network make balancing the rates of change much harder to maintain dissipativity in response to changes in loadability. 
\par A much simpler solution would be to place a controllable generation source at the load bus experiencing the changes in load. The dynamical limits in (\ref{powerTransBound}) include both active and reactive terms, a change in the real power of the load may be counterbalanced by an equal and opposite source of reactive power at the load side, such as synchronous condensers, inverters operating in grid following mode, or other novel approaches for providing reactive power on the load end, such as those presented in \cite{abessi2015centralized}. 
\par Suppose a source of reactive power is placed at a load bus providing an amount of reactive power less than the demanded reactive power by the load. The majority of the energy provided by the source of reactive power would be consumed by the load, with very little being transmitted into the transmission network. This additional source of reactive power will only contribute to the terms relating to dynamics of the generators and loads, as the behavior of the generator may change the dynamics of the load. The change in dynamic loadability is then:
\begin{equation}
    \Delta \sum_{k \in L} \left( \frac{1}{\tau^{tl}} +\frac{1}{\tau_k^l} \right) P_k^l \leq \sum_{k \in L} \Delta \dot{P}_k^l + \Delta \dot{Q}_k^l + \sum_{j \in G} \Delta \dot{P}_j^g + \Delta \dot{Q}_j^g
    \label{loadabilitychange}
\end{equation}

\section{SIMULATION RESULTS} \label{simresultssection}
\subsection{Dynamic Instabilities in PV Curves}
\begin{figure*}
    \centering
    \includegraphics[width=\textwidth]{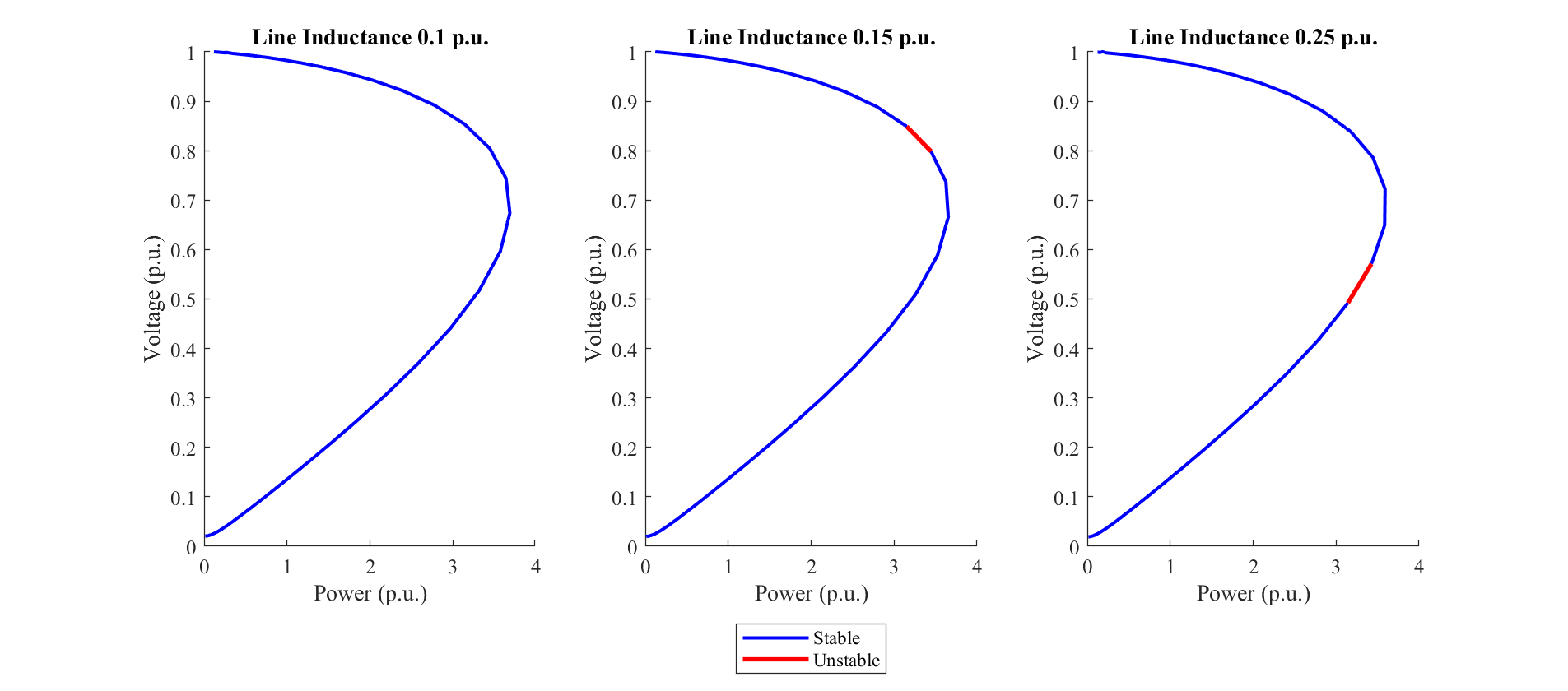}
    \caption{PV curves for the single generator, transmission line load system with three different values for the line inductance. Sections of the PV curve in blue represent regions where the operation was stable, and sections in red represent regions where the operation was unstable.}
    \label{PVCurves}
\end{figure*}

The system in Fig. \ref{fig:GTLL} was simulated using a full seven state model of the generator dynamics along with its governor and exciter control \cite{Cvetkovic2018}, a long line model of a transmission line using a lumped parameter equivalent $\pi$-model, and an RL load. The inductance value of the load was held constant while the resistor values were varied from the open circuit case to the short circuit one. The resultant PV curves are shown in Fig. \ref{PVCurves}.
\par Stability regions of the PV curves are determined by placing a small disturbance on the source, and the dynamical simulation is performed. The time constant of the line is inversely proportional to the line inductance. As expected by the maximum dynamic loadability limit, a weaker line will experience stability issues. This is demonstrated in the PV curves generated in Fig. \ref{PVCurves}. For the strong line with a line inductance of $0.1$ p.u., operation is stable over the entire curve. However, when the line weakens, in the $0.15$ p.u. and $0.25$ p.u. cases, there are regions of instability. This region appears in the high voltage solutions for the $0.15$ p.u. inductance line, and in the low voltage solution of the $0.25$ p.u. inductance line.
\subsection{Load Side Reactive Power Support}
\begin{figure}
    \centering

\tikzset{every picture/.style={line width=0.75pt}} 

\begin{tikzpicture}[x=0.75pt,y=0.75pt,yscale=-1,xscale=1]

\draw   (110,115) .. controls (110,101.19) and (121.19,90) .. (135,90) .. controls (148.81,90) and (160,101.19) .. (160,115) .. controls (160,128.81) and (148.81,140) .. (135,140) .. controls (121.19,140) and (110,128.81) .. (110,115) -- cycle ;
\draw    (160,115) -- (210,115) ;
\draw   (210,95.25) -- (280,95.25) -- (280,135.25) -- (210,135.25) -- cycle ;
\draw    (280.5,129.5) -- (306,129.64) -- (370,130) ;
\draw    (370,130) -- (370,148) ;
\draw [shift={(370,150)}, rotate = 270] [color={rgb, 255:red, 0; green, 0; blue, 0 }  ][line width=0.75]    (10.93,-3.29) .. controls (6.95,-1.4) and (3.31,-0.3) .. (0,0) .. controls (3.31,0.3) and (6.95,1.4) .. (10.93,3.29)   ;
\draw  [dash pattern={on 4.5pt off 4.5pt}] (106,80) -- (175,80) -- (175,149.5) -- (106,149.5) -- cycle ;
\draw  [dash pattern={on 4.5pt off 4.5pt}] (200,80.5) -- (290,80.5) -- (290,150) -- (200,150) -- cycle ;
\draw  [dash pattern={on 4.5pt off 4.5pt}] (320,120) -- (400,120) -- (400,150) -- (320,150) -- cycle ;
\draw    (190,70) -- (162,70) ;
\draw [shift={(160,70)}, rotate = 360] [color={rgb, 255:red, 0; green, 0; blue, 0 }  ][line width=0.75]    (10.93,-3.29) .. controls (6.95,-1.4) and (3.31,-0.3) .. (0,0) .. controls (3.31,0.3) and (6.95,1.4) .. (10.93,3.29)   ;
\draw    (180,160) -- (208,160) ;
\draw [shift={(210,160)}, rotate = 180] [color={rgb, 255:red, 0; green, 0; blue, 0 }  ][line width=0.75]    (10.93,-3.29) .. controls (6.95,-1.4) and (3.31,-0.3) .. (0,0) .. controls (3.31,0.3) and (6.95,1.4) .. (10.93,3.29)   ;
\draw    (310,160) -- (282,160) ;
\draw [shift={(280,160)}, rotate = 360] [color={rgb, 255:red, 0; green, 0; blue, 0 }  ][line width=0.75]    (10.93,-3.29) .. controls (6.95,-1.4) and (3.31,-0.3) .. (0,0) .. controls (3.31,0.3) and (6.95,1.4) .. (10.93,3.29)   ;
\draw    (340,160) -- (368,160) ;
\draw [shift={(370,160)}, rotate = 180] [color={rgb, 255:red, 0; green, 0; blue, 0 }  ][line width=0.75]    (10.93,-3.29) .. controls (6.95,-1.4) and (3.31,-0.3) .. (0,0) .. controls (3.31,0.3) and (6.95,1.4) .. (10.93,3.29)   ;
\draw    (280,99.5) -- (306,99.64) -- (365.5,99.5) ;
\draw    (320,80) -- (320,52) ;
\draw [shift={(320,50)}, rotate = 90] [color={rgb, 255:red, 0; green, 0; blue, 0 }  ][line width=0.75]    (10.93,-3.29) .. controls (6.95,-1.4) and (3.31,-0.3) .. (0,0) .. controls (3.31,0.3) and (6.95,1.4) .. (10.93,3.29)   ;
\draw   (340,55) .. controls (340,41.19) and (351.19,30) .. (365,30) .. controls (378.81,30) and (390,41.19) .. (390,55) .. controls (390,68.81) and (378.81,80) .. (365,80) .. controls (351.19,80) and (340,68.81) .. (340,55) -- cycle ;
\draw    (365,80) -- (365.5,99.5) ;
\draw  [dash pattern={on 4.5pt off 4.5pt}] (327.5,24) -- (396.5,24) -- (396.5,93.5) -- (327.5,93.5) -- cycle ;

\draw (125,106.9) node [anchor=north west][inner sep=0.75pt]    {$G_{1}$};
\draw (231.5,106.9) node [anchor=north west][inner sep=0.75pt]    {$TL_{1}$};
\draw (376,130.9) node [anchor=north west][inner sep=0.75pt]    {$L_{1}$};
\draw (141,42.4) node [anchor=north west][inner sep=0.75pt]    {$P_{1}^{g} ,\ \dot{Q}_{1}^{g}$};
\draw (151,164.4) node [anchor=north west][inner sep=0.75pt]    {$P_{1,1}^{tl} ,\ \dot{Q}_{1,1}^{tl}$};
\draw (261,163.4) node [anchor=north west][inner sep=0.75pt]    {$P_{1,2}^{tl} ,\dot{Q}{_{1,2}^{tl}}$};
\draw (337,164.4) node [anchor=north west][inner sep=0.75pt]    {$P_{1}^{l} ,\ \dot{Q}_{1}^{l}$};
\draw (291,52.4) node [anchor=north west][inner sep=0.75pt]    {$\dot{Q}_{2}^{g}$};
\draw (355,46.9) node [anchor=north west][inner sep=0.75pt]    {$G_{2}$};

\end{tikzpicture}
    \caption{Modified generator, transmission line, load system with an additional generator connected on the load side. This generator is configured to only provide reactive power.}
    \label{GTLLQsup}
\end{figure}
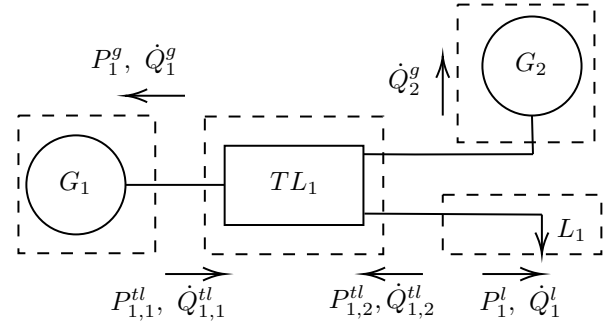

The system in Fig. \ref{fig:GTLL} was modified to add an additional generation source on the load side, though the generation source is configured to only provide reactive power. This modified system is shown in Fig. \ref{GTLLQsup}. The second generator, $G_2$, is configured as a negative PQ load, emblematic of an inverter in grid following mode, providing $P_0 = 0$ p.u., and $Q_0 = -Q_2^g$ p.u., for some predetermined set point. Following the perturbation of the system, the second generator is enabled, and its controls are given by (\ref{PQsourcedynamics}).
\begin{subequations}
\label{PQsourcedynamics}
    \begin{align}
        \frac{dP}{dt} &= \frac{-1}{\tau_P} (P - P_0) \\
        \frac{dQ}{dt} &= \frac{-1}{\tau_Q} (Q - Q_0)
    \end{align}
\end{subequations}
Where $\tau_P$ and $\tau_Q$ represent the time constants of tracking the set points, and $P$ and $Q$ denote the current output of the device as seen at its ports. 
\par
\begin{figure*}
    \centering
    \includegraphics[width=\linewidth]{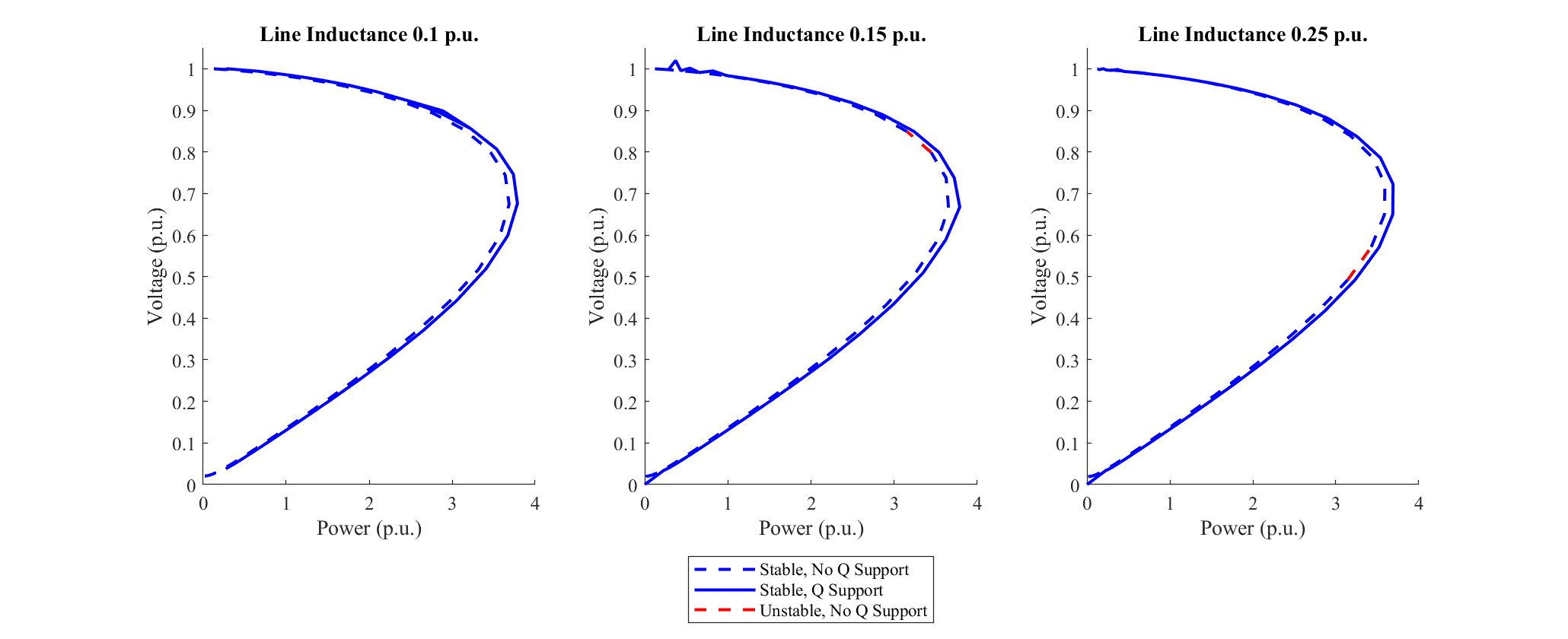}
    \caption{PV curves for the modified system with reactive power support on the load end. The entire range of the PV curve becomes stable with the added dynamic reactive power source, and also has higher maximum power deliverability.}
    \label{PVcurvesQsup}
\end{figure*}

 The same procedure was utilized as generating the PV curves in Fig. \ref{PVCurves}, except with this new generating unit connected at the load side. The system remained unstable for time constant $\tau_P = \tau_Q = .1$. When the time constant was further reduced to $\tau_P = \tau_Q = .01$, the system stabilized for the entire range of loads. The set point of $Q_2^g = 0.1$ p.u. was selected to be less than the reactive power demand of the load. The PV curves of the modified system are shown overlaid on those of the unmodified system in Fig. \ref{PVcurvesQsup}. Comparing the maximum power point of each curve, the maximum power of the system with the dynamic reactive power support at the load end is larger. This comparison is summarized in Table \ref{maxpowerpoint}.
\par The set point of the reactive power support is configured to be less than the reactive power demand by the load. Computing an upper bound on the change to the system loadability as given in (\ref{loadabilitychange}) gives $0.1$ p.u., which shows good agreement with the changes noted in Table \ref{maxpowerpoint}.

\begin{table}
    \centering
    \caption{Maximum Dynamic Loadability With and Without Reactive Power Support}
    \begin{tabular}{ccccc}
    \hline \hline
        \multirow{2}{*}{\textbf{Line Inductance (p.u)}} & \multicolumn{2}{c}{Max. Power Point (p.u.)} & \\ \cline {2-3}
         & No $Q$ Support & With $Q$ Support & \\ \hline
        0.1 & 3.69 & 3.78 & \\
        0.15 & 3.65 & 3.79 & \\ 
        025 & 3.59 & 3.69 &  \\ \hline \hline
    \end{tabular}
    \label{maxpowerpoint}
\end{table}

\begin{figure}
    \centering
    \includegraphics[width = \columnwidth]{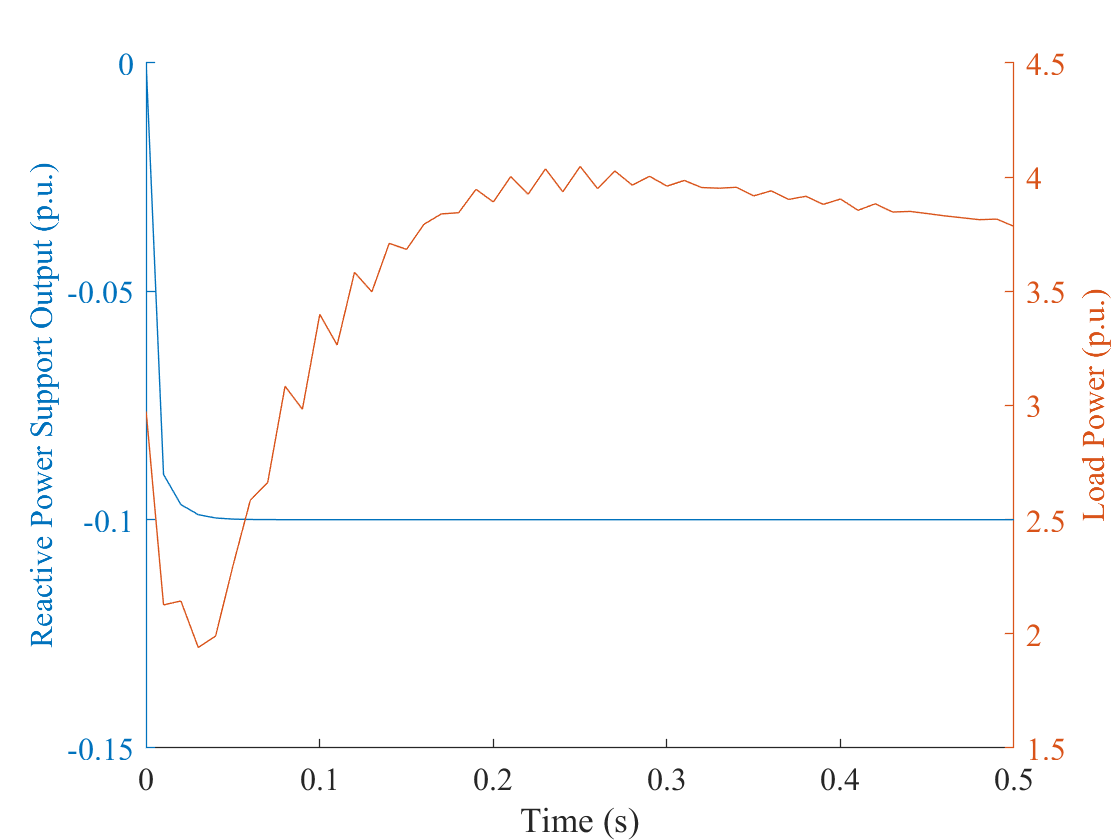}
    \caption{Time response of the RL load following a small disturbance and the dynamics of the generating unit at the load side. }
    \label{fig:enter-label}
\end{figure}

\section{Conclusions} \label{conclusion}
We have presented a method to use a general energy dynamics modeling framework to establish sufficient conditions on the loadability of a transmission network, given the dynamics of generators, lines, and loads. These conditions show the key role that not only the level of generation plays, but also the rates at which generation and load change, as well as the internal properties of the transmission system. These conditions were used to motivate the role of load side reactive power supportin system stability by providing very fast changing reactive power levels in response to a system disturbance. The proposed technique is entirely technology agnostic, and could be applied to any arbitrary network configuration, comprised of a mix of generation sources and load types.


\section{Acknowledgements} \label{acknowledgements}
The authors greatly appreciate conversations and insights provided by Dr. Rupamathi Jaddivada and Dr. Xia Miao which greatly aided this manuscript.  

\bibliographystyle{IEEEtran}

\end{document}